\documentclass[aps,prd,twocolumn,groupedaddress,amssymb,eqsecnum,showpacs,epsfig]{revtex4}
\usepackage{graphicx}
\usepackage{bm}
\usepackage{dcolumn}
\usepackage{amsmath}
\def \lleq {\lower0.9ex\hbox{ $\buildrel < \over \sim$} ~}
\def \ggeq {\lower0.9ex\hbox{ $\buildrel > \over \sim$} ~}

\def \beq  {\begin{equation}}
\def \eeq  {\end{equation}}
\def \ber  {\begin{eqnarray}}
\def \eer  {\end{eqnarray}}

\def\apj{{Astroph.\@ J.\ }}

\def\aj{{Astron.\@ J.\ }}

\begin{document}
\newcommand{\newc}{\newcommand}

\newc{\be}{\begin{equation}}
\newc{\ee}{\end{equation}}
\newc{\ba}{\begin{eqnarray}}
\newc{\ea}{\end{eqnarray}}
\newc{\bea}{\begin{eqnarray*}}
\newc{\eea}{\end{eqnarray*}}
\newc{\D}{\partial}
\newc{\ie}{{\it i.e.} }
\newc{\eg}{{\it e.g.} }
\newc{\etc}{{\it etc.} }
\newc{\etal}{{\it et al.}}
\newcommand{\nn}{\nonumber}
\newc{\ra}{\rightarrow}
\newc{\lra}{\leftrightarrow}
\newc{\lsim}{\buildrel{<}\over{\sim}}
\newc{\gsim}{\buildrel{>}\over{\sim}}
\title{Evolving Newton's Constant, Extended Gravity Theories and SnIa Data Analysis}
\author{S. Nesseris and L. Perivolaropoulos}
 \affiliation{Department of
Physics, University of Ioannina, Greece}
\date{\today}

\begin{abstract}
If Newton's constant $G$ evolves on cosmological timescales as
predicted by extended gravity theories then Type Ia supernovae
(SnIa) can not be treated as standard candles. The
magnitude-redshift datasets however can still be useful. They can
be used to simultaneously fit for both $H(z)$ and $G(z)$ (so that
local $G(z)$ constraints are  also satisfied) in the context of
appropriate parameterizations. Here we demonstrate how can this
analysis be done by applying it to the Gold SnIa dataset. We
compare the derived effective equation of state parameter $w(z)$
at best fit with the corresponding result obtained by neglecting
the evolution $G(z)$. We show that even though the results clearly
differ from each other, in both cases the best fit $w(z)$ crosses
the phantom divide $w=-1$. We then attempt to reconstruct a scalar
tensor theory that predicts the derived best fit forms of $H(z)$
and $G(z)$. Since the best fit $G(z)$ fixes the scalar tensor
potential evolution $F(z)$, there is no ambiguity in the
reconstruction and the potential $U(z)$ can be derived uniquely.
The particular reconstructed scalar tensor theory however,
involves a change of sign of the kinetic term $\Phi'(z)^2$ as in
the minimally coupled case.
\end{abstract}
\pacs{98.80.Es,98.65.Dx,98.62.Sb}
\maketitle

\section{Introduction}
A diverse set of cosmological observations including the abundance
of galaxy clusters \cite{lss}, the baryon fraction in galaxy
clusters \cite{bfrac},  statistics of large scale redshift surveys
\cite{lsstat}, and the angular power spectrum of the Cosmic
Microwave Background (CMB) \cite{cmb} indicate that the universe
is flat and that there is a low value of the matter density
parameter $0.16<\Omega_{0m}<0.35$. Thus, in the context of
standard general relativistic cosmology there is a gap  between
$\Omega_{0m}$ and $\Omega_{tot}=1$ required for flatness. This gap
is usually assumed to be filled by an unknown form of energy
called {\it dark energy}\cite{dark energy}. In addition, the
magnitude-redshift relation for type Ia supernovae (SnIa)
\cite{snobs,Riess:2004nr,snls}, which can probe the recent
expansion history of the universe indicates that the universe has
entered a phase of accelerating expansion (the scale factor obeys
${\ddot a}>0$). This can be reconciled with the other cosmological
data by assuming that the dark energy has negative pressure and
therefore has repulsive gravitational properties (see
\cite{Perivolaropoulos:2006ce} for a recent review).

The dark energy component is usually described by an equation of
state parameter $w\equiv{p\over \rho}$ (the ratio of the
homogeneous dark energy pressure $p$ over the energy density
$\rho$). For cosmic acceleration, a value of $w<-{1\over 3}$ is
required as indicated by the Friedmann equation \be {{\ddot
a}\over a}=-{{4\pi G}\over 3}(\rho +3p) \label{fried}\ee The
simplest viable example of dark energy is the cosmological
constant ($w=-1$). This example however even though consistent
with present data lacks physical motivation. Questions like `What
is the origin of the cosmological constant?' or `Why is the
cosmological constant $10^{120}$ times smaller than its natural
scale $M_{pl}$ so that it starts dominating at recent cosmological
times (coincidence problem)?' remain unanswered. Attempts to
replace the cosmological constant by a dynamical scalar field
(quintessence\cite{quintess}) which may also couple to dark matter
\cite{coupdm}, have created a new problem regarding the initial
conditions of quintessence which even though can be resolved in
particular cases (tracker quintessence), can not answer the above
questions in a satisfactory way.

An alternative approach towards understanding the nature of dark
energy is to attribute it to extensions of general
relativity\cite{modgrav} on cosmological scales. Such extensions
can be expressed for example through scalar-tensor
theories\cite{Esposito-Farese:2000ij}. In these theories the
Einstein Lagrangian of general relativity is replaced by a
generalized Lagrangian of the form \be {\cal
L}=\frac{F(\Phi)}{2}~R - \frac{Z(\Phi)}{2}~g^{\mu\nu}
\partial_{\mu}\Phi
\partial_{\nu}\Phi
- U(\Phi)  + {\cal L}_m[\psi_m; g_{\mu\nu}]\  \label{ljf} \ee
where we have set $8\pi G=1$ ($F_0=1$) and ${\cal L}_m$ represents
the matter fields and does not depend on $\Phi$ so that the weak
equivalence principle is satisfied. A common choice in the Jordan
frame is to set $Z\rightarrow 1$ by rescaling the field $\Phi$
\cite{Esposito-Farese:2000ij}.

The evolution of Newton's constant predicted in the context of
extended gravity theories requires special care when comparing the
predictions of these theories with
observations\cite{Acquaviva:2004ti}. This evolution induces
special effects to the physics of SnIa
\cite{Amendola:1999vu,Garcia-Berro:1999bq,Gaztanaga:2001fh,Riazuelo:2001mg}.
The observed magnitude redshift relation of SnIa can be translated
to luminosity distance-redshift relation (which leads to the
expansion history $H(z)$) only under the assumption that SnIa
behave as standard candles. This assumption is justified in view
of the fact that the observational light curves of closeby SnIa
are well understood and their individual intrinsic differences can
be accounted for. Nevertheless, since the accelerated expansion of
the universe is based on the fact that the peak luminosities of
distant supernovae appear to be $\sim 0.20$ magnitude fainter than
predicted for an empty universe and $\sim 0.25$ magnitude fainter
than a decelerating universe with $\Omega_{0m}=0.3$, it is clear
that even minor unaccounted evolutionary effects can drastically
change our current view for accelerating universe. The possible
consequences of evolutionary effects in SnIa due to changes in the
zero age mass and metallicity of the progenitor star have been
previously explored \cite{Dominguez:1998jt} who found that changes
in the underlying population cause a change in the maximum
brightness by about $0.1-0.2$ magnitudes.

The corresponding evolutionary effects due to the evolution of $G$
in scalar tensor theories have also been studied
\cite{Riazuelo:2001mg,Garcia-Berro:1999bq}. The peak luminosity of
SnIa is proportional \cite{Gaztanaga:2001fh} to the mass of nickel
synthesized which is a fixed fraction of the Chandrasekhar mass
$M_{Ch}$ varying as $M_{Ch}\sim G^{-3/2}$. Therefore the SnIa peak
luminosity varies like $L\sim G^{-3/2}$ and the corresponding SnIa
absolute magnitude evolves like \be M-M_0=\frac{15}{4}
log\frac{G}{G_0} \label{abmag} \ee where the subscript $0$ denotes
the local values of $M$ and $G$. Thus, the magnitude-redshift
relation of SnIa in the context of extended gravity theories is
connected with the luminosity distance $d_L (z)$ as \be m_{th}(z)
= M_0 + 5 log d_L (z) + \frac{15}{4} log\frac{G(z)}{G_0}
\label{magred} \ee In the limit of constant $G$ this reduces to
the familiar result. On the other hand, in scalar tensor
theories\cite{Esposito-Farese:2000ij} we have \be \frac{G(z)}{G_0}
= \frac{1}{F} \frac{2F +  4 (d F/d \Phi)^2}{2F + 3 (d F/d
\Phi)^2}\simeq \frac{1}{F} \label{gf} \ee  and solar system
experiments \cite{Will:2005va,will-bounds} indicate that
$\frac{dF(\Phi)}{d \Phi} \sim \frac{dF(z)}{d z}\simeq 0$. Assuming
flatness, the expansion history $H(z)$ is obtained from \be d_L
(z) = (1+z)\int_0^z dz' \sqrt{\frac{G_0}{G(z')}} \frac{1}{H(z')}
\label{dlh} \ee Therefore, by fitting $m_{th}$ of (\ref{magred})
to the observed Gold SnIa \cite{Riess:2004nr} dataset expressed as
$m_{obs}(z_i)$ and using (\ref{gf}) and (\ref{dlh}) we may obtain
the best fit forms of both $H(z)$ and $G(z)$ assuming appropriate
parameterizations.  It should be pointed out that even though the
modified magnitude-redshift relation (\ref{magred}) has been known
for some time \cite{Garcia-Berro:1999bq}, it has not been properly
utilized in studies attempting to constrain extended theories of
gravity with SnIa data (see eg \cite{Caresia:2003ze}). This task
is undertaken in what follows.

The structure of this paper is the following: In the next section
we use simple polynomial parameterizations of $G(z)$ and $H(z)$ to
fit these functions to the Gold dataset using equations
(\ref{magred}) and (\ref{dlh}). In section III we use the best
fits of $G(z)$ and $H(z)$ to construct the best fit scalar tensor
theory ie the potential $U(z)$ and the form of $\Phi'(z)^2$.
Finally in section IV we conclude, summarize and discuss possible
implications of our results.

\section{Fitting $G(z)$ and $H(z)$ to the Gold dataset}

The recent expansion history of the universe is best probed by
using a diverse set of cosmological data including SnIa standard
candles, CMB spectrum, large scale structure power spectrum, weak
lensing surveys etc. Given the present quality of cosmological
data, among the above cosmological observations the most sensitive
and high quality probe of the recent expansion history $H(z)$ is
the magnitude-redshift relation of SnIa. This probe will be used
in the present study.

There have been two approaches in deriving\cite{phant-obs2} the
functions $d_L(z)$, $H(z)$ and $w(z)$ from the discrete set of
magnitude-redshift data and their errorbars. According to the
first approach\cite{Shafieloo:2005nd}, a smoothing window function
is used to derive a continuous function $m(z)$ (or equivalently
$d_L(z)$) assuming no evolution of $G$ in equation (\ref{magred}))
and then $H(z)$ and $w(z)$ are obtained using the well known
relations \cite{Sahni:2004ai} \be \label{hz1} H(z)= c [{d\over
{dz}} ({{d_L(z)}\over {1+z}})]^{-1} \ee and \be \label{wz3}
w(z)={{p_{DE}(z)}\over {\rho_{DE}(z)}}={{{2\over 3} (1+z) {{d \ln
H}\over {dz}}-1} \over {1-({{H_0}\over H})^2 \Omega_{0m} (1+z)^3}}
\ee This approach works well in the context of general relativity
where $G(z)=G_0$ but it can not be used in the context of
scalar-tensor theories (unless supplemented by additional
cosmological observations) because equation (\ref{magred}) alone
can not be used to derive both $d_L(z)$ and $G(z)$ from the single
smoothed function $m(z)$. In fact, even if a smoothed form of
$d_L(z)$ was obtained it would be hard to disentangle $H(z)$ from
$G(z)$ in equation (\ref{dlh}).

The second approach\cite{alam1,nesper,Lazkoz:2005sp,Weller:2001gf}
is based on assigning particular parameterizations (which are
consistent with other observations) to the function $H(z)$ (and to
$G(z)$ if applicable) involving two to three parameters  and
fitting these parameters to the magnitude-redshift data. The
goodness of fit corresponding to any set of parameters
$a_1,...,a_n$ is determined by the probability distribution of
$a_1,...,a_n$ \ie \be P({\bar M}, a_1,...,a_n)= {\cal N} e^{-
\chi^2({\bar M},a_1,...,a_n)/2} \label{prob1} \ee where  \be
\chi^2 ({\bar M},a_1,...,a_n)= \sum_{i=1}^{157}
\frac{(m^{obs}(z_i) - m^{th}(z_i;{\bar
M},a_1,...,a_n))^2}{\sigma_{m^{obs}(z_i)}^2} \label{chi2} \ee and
${\cal N}$ is a normalization factor. If prior information is
known on some of the parameters $a_1,...,a_n$ then we can either
fix the known parameters using the prior information or
`marginalize', i.e. average the probability distribution
(\ref{prob1}) around the known value of the parameters with an
appropriate `prior' probability distribution.

The parameters ${\bar a}_1,...,{\bar a}_n$ that minimize the
$\chi^2$ expression (\ref{chi2}) are the most probable parameter
values (the `best fit') and the corresponding $\chi^2({\bar
a}_1,...,{\bar a}_n)\equiv \chi_{min}^2$ gives an indication of
the quality of fit for the given parametrization: the smaller
$\chi_{min}^2$ the better the parametrization. The minimization
with respect to the parameter $\bar{M}$ can be made trivially by
expanding the $\chi^2$ of equation (\ref{chi2}) with respect to
$\bar{M}$ as\cite{Lazkoz:2005sp} \be  \chi^2 (a_1,..,a_n) = A - 2
{\bar M} B  + {\bar M}^2 C  \label{chi2bm} \ee where \ba
A(a_1,..,a_n)&=&\sum_{i=1}^{157} \frac{(m^{obs}(z_i) - m^{th}(z_i
;{\bar
M}=0,a_1,..,a_n))^2}{\sigma_{m^{obs}(z_i)}^2} \label{bb} \nn \\
B(a_1,..,a_n)&=&\sum_{i=1}^{157} \frac{(m^{obs}(z_i) - m^{th}(z_i
;{\bar
M}=0,a_1,..,a_n))}{\sigma_{m^{obs}(z_i)}^2} \label{bb} \nn \\
C&=&\sum_{i=1}^{157}\frac{1}{\sigma_{m^{obs}(z_i)}^2 } \label{cc}
\ea Equation (\ref{chi2bm}) has a minimum for ${\bar M}={B}/{C}$
at \be \chi'^2(a_1,...,a_n)=A(a_1,...,a_n)-
\frac{B(a_1,...,a_n)^2}{C} \label{chi2min1}\ee Thus instead of
minimizing $\chi^2({\bar M},a_1,...,a_n)$ we can minimize
$\chi'^2(a_1,...,a_n)$ which is independent of ${\bar M}$.
Obviously $\chi_{min}^2=\chi_{min}^{'2}$.

The errors are evaluated using the covariance
matrix of the fitted parameters \cite{press92} %%%%
and the error on any cosmological quantity, \eg the equation of
state $w(z;p_i)$, is given by: \be \sigma_w^2
=\sum_{i=1}^n(\frac{\partial w}{\partial p_i})C_{ii}+
2\sum_{i=1}^n\sum_{j=i+1}^n (\frac{\partial w}{\partial p_i})
(\frac{\partial w}{\partial p_j}) C_{ij} \ee where $p_i$ are the
cosmological parameters and $C_{ij} $ the covariance matrix
\cite{Alam:2004ip}.

We considered simple polynomial parameterizations for the
functions $H(z)$ and $G(z)$ of the form \ba H^2(z) &=&  H_0^2 \{
\Omega_{0m}
(1+z)^3 + a_1(1+z)+a_2(1+z)^2+ \nn \\ &&(1-\Omega_{0m}-a_1-a_2) \} \label{hzanz} \\
G(z)&=& G_0 \; (1+a\; z^2)\label{gzanz} \ea
\begin{figure}[h]
\centering
\includegraphics[bb=80 400 380 740,width=6.7cm,height=7cm,angle=0]{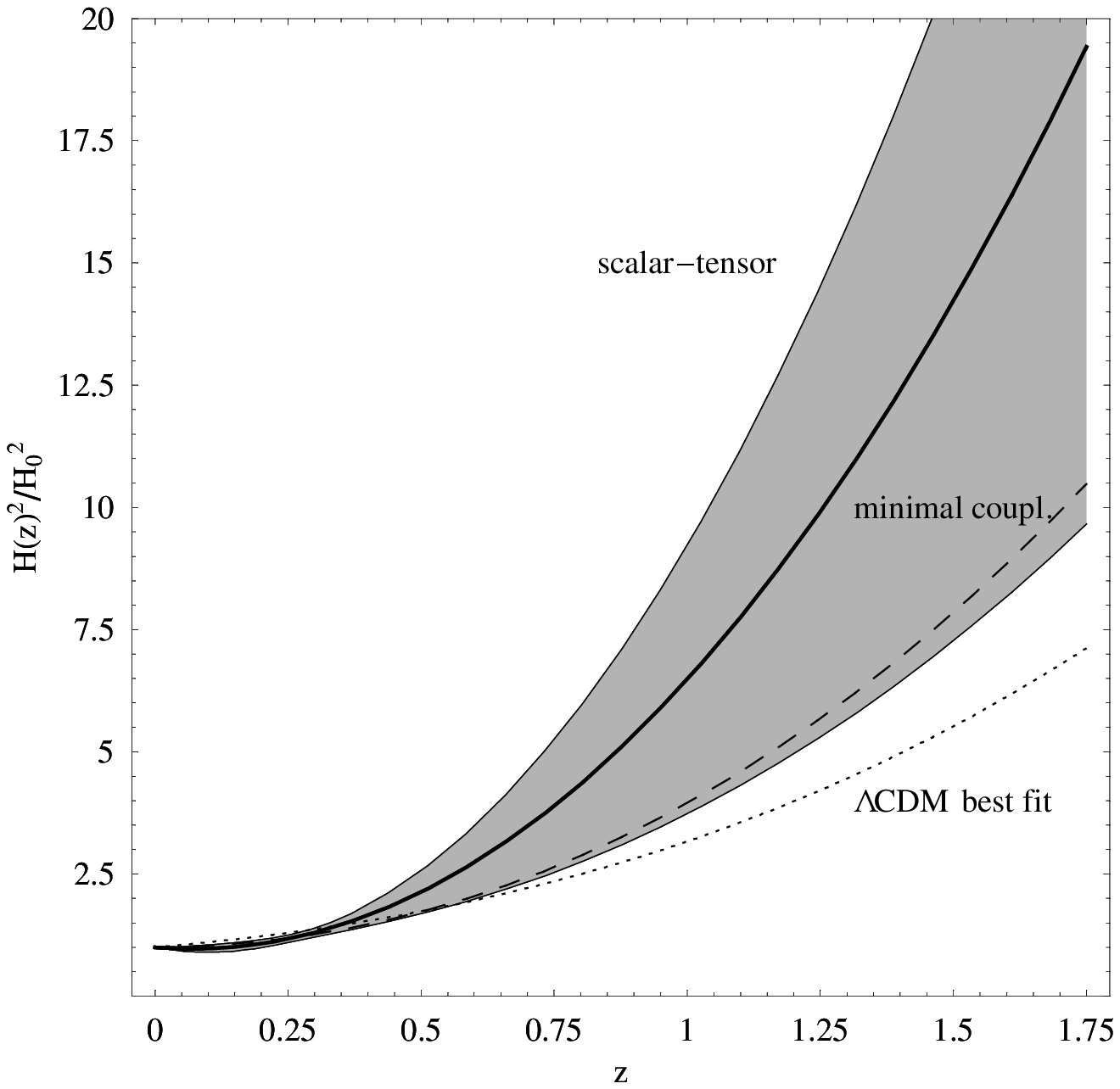}
\caption{The best fit form of $H(z)$ in the scalar tensor
(continuous line), minimally coupled (dashed line) and
$\Lambda$CDM (dotted) cases. The shaded region corresponds to the
$1\sigma$ region of the scalar-tensor best fit.} \label{fig1}
\end{figure}

\noindent where the linear term in $G(z)$ has been ignored due to
experimental constraints on scalar tensor theories
\cite{will-bounds}. Using the parameterizations (\ref{hzanz}) and
(\ref{gzanz}) we have minimized the $\chi'^2$ of (\ref{chi2min1})
using the 157 datapoints of the Gold dataset and equations
(\ref{magred}), (\ref{dlh}). We used a prior of $\Omega_{0m}=0.23$
\cite{wmap3}  and we verified that our results are relatively
insensitive to the prior of $\Omega_{0m}$ used in the range
$0.18<\Omega_{0m}<0.32$. The minimum was obtained at
$\chi^2=173.045$ for $a_1=-12.35\pm 8.55 $, $a_2=5.41\pm 3.82  $
and $a=0.05 \pm 0.04$. We compared our results with the
corresponding minimally coupled case obtained by fixing $a=0$
before minimization (setting $G(z)=G_0$ at all times). The
corresponding minimum was obtained at $\chi^2=174.168$ for
$a_1=-4.54\pm 2.52$, $a_2=1.96\pm 1.09$.

The best fit functions for $H(z)$, $w(z)$ and
$F(z)=\frac{1}{G(z)}$ for both the scalar-tensor and minimally
coupled cases are shown in Figs. 1, 2 and 3 respectively along
with the $1\sigma$ (shaded) region of the scalar-tensor best fit.

An interesting feature of Fig. 3 is the relatively large (about
15$\%$) cosmological variation of Newton's constant. In
 \cite{Umezu:2005ee} it is shown that CMB and SDSS
constraints allow Newton's constant to vary by a factor of 2 on
cosmological scales. The variation implied by our analysis is
about $15\%$ and is well within the cosmological constraints.

Also, it is clear that the best fit equation of state parameter
$w(z)$ crosses the phantom divide in both the scalar-tensor and
the minimally coupled case at about $z\simeq 0.2$. This type of
crossing which seems to be favored by the Gold SnIa dataset
\cite{crossgold,alam1,nesper,Lazkoz:2005sp} (but not
\cite{nocrossnls,nocrossnls1} by the more recent first year SNLS
dataset \cite{snls}) has been the subject of extensive studies in
the literature\cite{crosstud} as its reproduction is highly
non-trivial in the context of most theoretical
\enlargethispage{\baselineskip} models\cite{Vikman:2004dc}.

\begin{figure}[h]
\centering
\includegraphics[bb=80 20 345 350,width=6.7cm,height=7cm,angle=0]{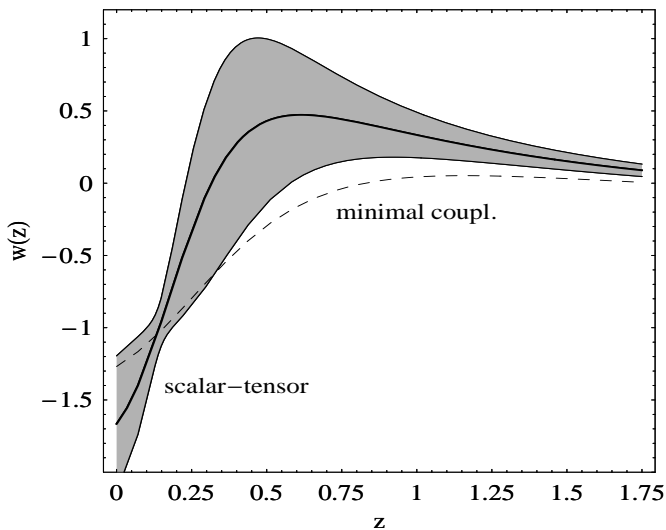}
\caption{The best fit form of $w(z)$ in the scalar tensor
(continuous line) and minimally coupled (dashed line) cases.}
\label{fig2}
\end{figure}

In the scalar-tensor case $w(z)$ does not have the usual meaning
of the dark energy equation of state but it is merely defined in
terms of $H(z)$ as in equation (\ref{wz3}). In particular we use
$w(z)$ simply as an alternative way of plotting $H(z)$. Such a way
is useful for comparing with other analyses
\cite{Sahni:2004ai,alam1} based on dark energy. Those analyses
also derive their best fit $w(z)$ from their best fit $H(z)$ but
they simply interpret the derived best fit $w(z)$ as an equation
of state parameter. Therefore even though the interpretation is
different, the comparison is meaningful since the actual quantity
that is compared is $H(z)$. The best fit functions $H(z)$ and
$F(z)$ will be used in the next section to complete the
construction of the best fit scalar tensor theory from the Gold
dataset.

\section{Reconstructing the Scalar-Tensor Lagrangian}
The derived best fit functions $F(z)$ and $H(z)$ may now be used
as input \cite{Perivolaropoulos:2005yv,Tsujikawa:2005ju} in the
field equations obtained from the Lagrangian (\ref{ljf}) in a
cosmological setup to obtain the potential $U(z)$ and the field
kinetic term $\Phi'(z)^2$.

Assuming a homogeneous $\Phi$ and varying the action corresponding
to (\ref{ljf}) in background of a flat FRW metric \be ds^2 = -dt^2
+ a^2(t)(dr^2 + r^2 \left(d\theta^2 + \sin^2\theta~d\phi^2\right))
\label{dl2}\ee we find the coupled system of
equations\cite{Esposito-Farese:2000ij}
\begin{eqnarray}
3F\cdot H^2 &=&  \rho +{1\over 2} \dot\Phi^2 - 3 H \cdot \dot F +
U
\label{fe1}\\
-2F\cdot\dot H  &=& (\rho+p) + \dot \Phi^2 +\ddot F - H\cdot \dot
F \label{fe2} \label{dmt}
\end{eqnarray}
where we have assumed the presence of a perfect fluid $(\rho,p)$.
Eliminating $\dot \Phi^2$ from (\ref{dmt}), setting \be q(z)\equiv
H(z)^2/H_0^2\label{qz} \ee and rescaling $U\rightarrow U\cdot
H_0^2$ while expressing in terms of redshift $z$ we obtain
\begin{widetext}
\ba F'' &+& \left[\frac{q'}{2q}-\frac{4}{1+z}\right]~F' +
\left[\frac{6}{(1+z)^2} - \frac{2}{(1+z)}\frac{q'}{2q}\right]~F =
\frac{2  U}{(1+z)^2 q^2} + 3 \frac{1+z}{q^2}
\Omega_{0m}\  \label{fe1a} \\
\Phi'^2 &=& -{6 F'\over 1+z} + {6 F\over (1+z)^2} -{{2 U}\over
(1+z)^2 q^2}  - 6 \frac{1+z}{q^2}  \Omega_{0m} \label{fe2a} \ea
\end{widetext}

\begin{figure}[h]
\centering
\includegraphics[bb=80 20 395 400,width=6.7cm,height=7cm,angle=0]{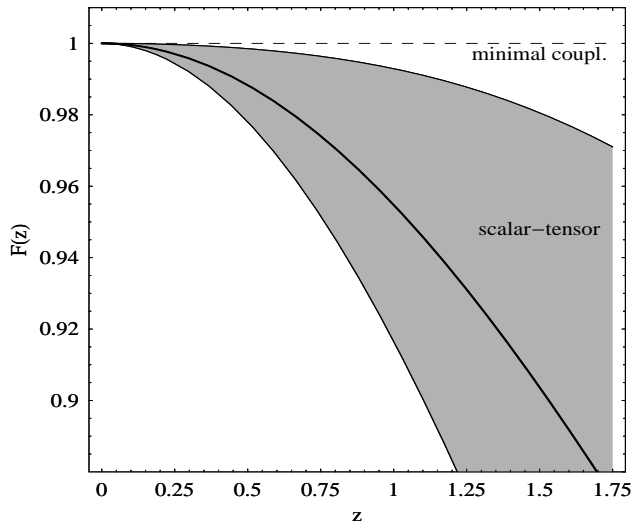}
\caption{The best fit form of $F(z)=1/G(z)$ in the scalar tensor
(continuous line) and minimally coupled (dashed line) cases. Due
to experimental constraints $F(z)$ displays an extremum at the
present time. Consistency with the SnIa data requires that this
extremum be a maximum for $F(z)$ (minimum for $G(z)$).}
\label{fig3}
\end{figure}

where the prime $'$ denotes differentiation with respect to
redshift $(\frac{d}{dz})$ and we have assigned properties of
matter ($p=0,\; \Omega_{0m}=\frac{3\rho_{0m}}{H_0^2}$) to the
perfect fluid.

Given the best fit form of $F(z)$ and $H(z)$ obtained in the
previous section from observations, equations (\ref{fe1a}) and
(\ref{fe2a}) may be used to reconstruct the $U(z)$ and
$\Phi'(z)^2$ which predict the best fit forms of $H(z)$ ($q(z)$)
and $F(z)$ for both the minimally coupled and the scalar-tensor
cases. The resulting forms of $U(z)$ and $\Phi'(z)^2$ are shown in
Figs 4 and 5 respectively.

\begin{figure}[h]
\centering
\includegraphics[bb=90 15 385 370,width=6.7cm,height=7cm,angle=0]{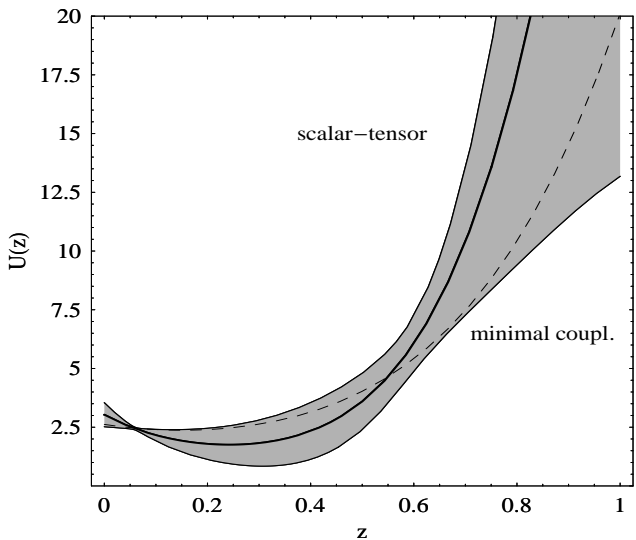}
\caption{The redshift evolution of the potential for the best fit
forms of $H(z)$ and $F(z)$ for the scalar-tensor (continuous line)
and minimally coupled (dashed line) cases.} \label{fig4}
\end{figure}
%30 100 470 700

\begin{figure}[h]
\centering
\includegraphics[bb=70 405 365 750,width=6.7cm,height=7cm,angle=0]{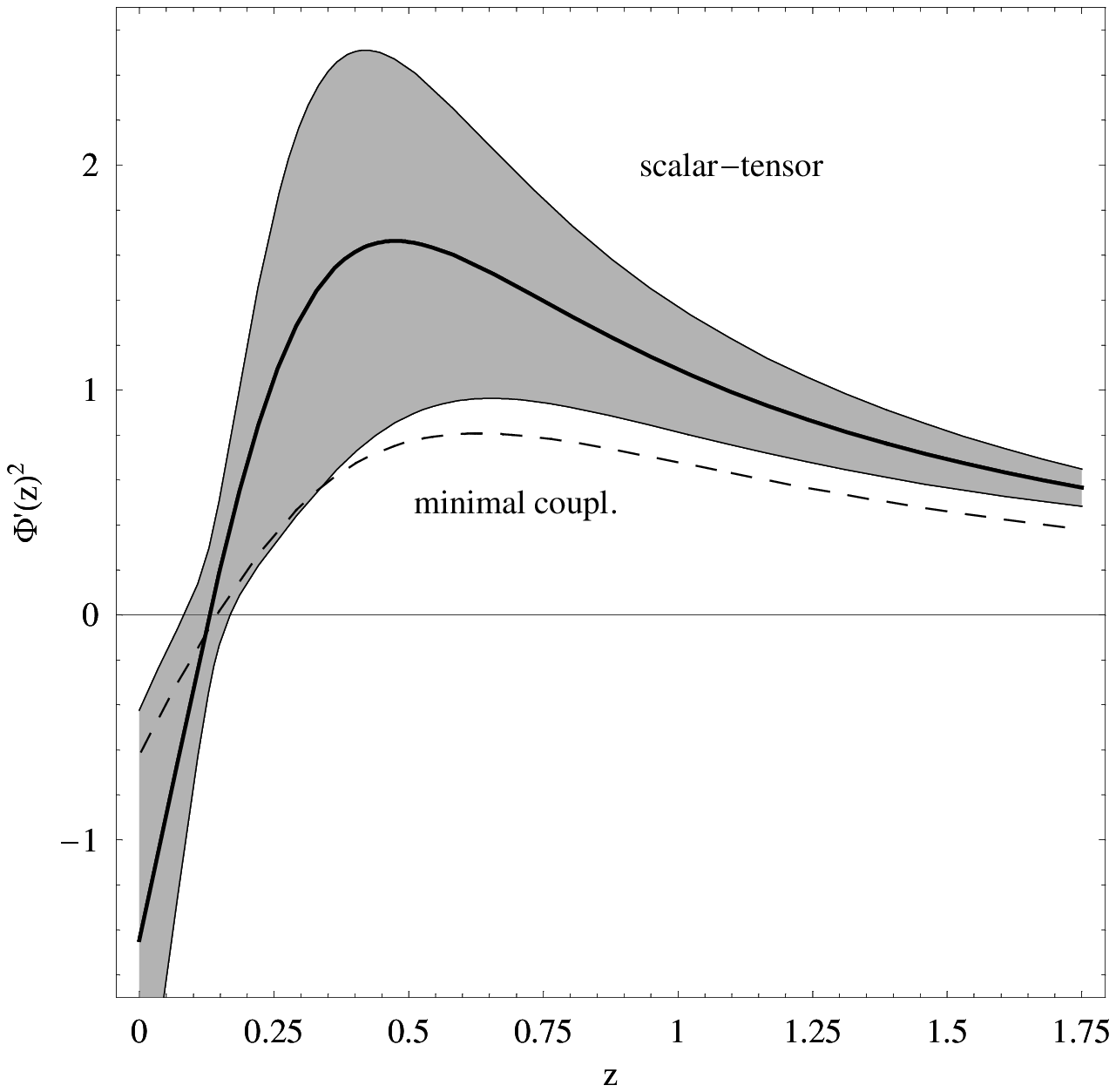}
\caption{The redshift evolution of the kinetic term $\Phi'(z)^2$
for the best fit forms of $H(z)$ and $F(z)$ for the scalar-tensor
(continuous line) and minimally coupled (dashed line) cases. The
change of sign at a redshift $z\simeq 0.2$ creates problems for
both classes of theories.} \label{fig4}
\end{figure}
An interesting feature of Fig. 5 is the change of sign of
$\Phi'^2(z)$ that is predicted for both classes of theories (at
best fit) at a redshift $z\simeq 0.2$ (the same redshift where the
phantom divide barrier is crossed). This creates a serious
challenge for both classes of theories. This problem was well
known for minimally coupled theories but it was believed that
scalar-tensor theories which have the potential to cross the
phantom divide in a self consistent way
\cite{Boisseau:2000pr,Perivolaropoulos:2005yv} could bypass this
problem. Our result however indicates that if the SnIa evolution
is taken into account then scalar-tensor theories are faced with a
similar problem as minimally coupled models\cite{Vikman:2004dc} in
the context of the Gold SnIa dataset.

As discussed in Ref. \cite{nocrossnls}, the Gold dataset mildly
favours dynamical dark energy compared to $\Lambda$CDM while the
SNLS dataset favours $\Lambda$CDM even in the context of dynamical
parameterizations. Thus we anticipate that in the case of the SNLS
dataset both of the best fit curves of Fig 2 (for the
scalar-tensor and minimally coupled cases) would tend to be closer
to the w=-1 line. In fact, the minimally coupled case with the
SNLS dataset has been discussed in Ref. \cite{nocrossnls} in the
context of the polynomial parameterizations of eq. (2.9).

\section{Conclusion}
We have constructed a method to utilize the magnitude redshift
SnIa data in the context of extended gravity theories. Assuming
simple redshift parameterizations for $H(z)$ and $G(z)$ we have
found their best fit forms and the corresponding error regions.
The best fit form of $G(z)$ indicates a slowly decreasing Newton's
constant (increasing $F(z)= \frac{1}{G(z)}$) at recent
cosmological times. The corresponding best fit form of $w(z)$
(defined through eq. (\ref{wz3})) was found to cross the phantom
divide $w=-1$ for both a constant and a redshift dependent $G$.
However, in the later case the best fit $w(z)$ was found to vary
more rapidly with redshift.

The simultaneous knowledge of both $F(z)$ and $H(z)$ allows the
unambiguous reconstruction of a scalar tensor theory by solving
the generalized Friedman equations. The particular reconstructed
scalar tensor theory using the Gold SnIa dataset and the specific
parameterizations, turned out to suffer from a similar problem as
the corresponding minimally coupled theory (the kinetic term
changes sign at a recent redshift). We have also considered more
complicated parameterizations for $G(z)$ and $H(z)$ which however
had only a minor effect on our results and did not seem to
alleviate the above mentioned problem.

Thus, if the best fit forms of $H(z)$ and $G(z)$ are verified by
future SnIa datasets, this would indicate that neither minimally
coupled nor extended quintessence are realized in Nature. In that
case the SnIa data could possibly be consistent with either
alternative extensions of general relativity (eg brane
worlds\cite{modgrav}) or by a combination of phantom +
quintessence scalars (quintom models\cite{Guo:2004fq}) (see also
\cite{Stefancic:2005cs,Caldwell:2005ai,Hu:2004kh} for alternative
approaches). An interesting extension of the present work would be
to use the best fit forms of $H(z)$ and $G(z)$ obtained from SnIa
and other cosmological data, in an attempt to reconstruct
consistently alternative extended gravity theories.

The Mathematica file with the numerical analysis of the paper can
be found at http://leandros.physics.uoi.gr/snevol.html .

{\bf Acknowledgements:} This research was funded by the program
PYTHAGORAS-1 of the Operational Program for Education and Initial
Vocational Training of the Hellenic Ministry of Education under
the  Community Support Framework and the European Social Fund. SN
acknowledges support from the Greek State Scholarships Foundation
(I.K.Y.).

%\widetext
\end{document}